# Spin-mediated charge-to-heat current conversion phenomena in ferromagnetic binary alloys

Asuka Miura[1], Ryo Iguchi[1], Takeshi Seki[1,2,3], Koki Takanashi[2,3,4], and Ken-ichi Uchida[1,2,3,a]

[1]National Institute for Materials Science, Tsukuba 305-0047, Japan

[2]Institute for Materials Research, Tohoku University, Sendai 980-8577, Japan

[3]Center for Spintronics Research Network, Tohoku University, Sendai 980-8577, Japan

[4]Center for Science and Innovation in Spintronics, Core Research Cluster, Tohoku University, Sendai 980-8577, Japan

[a]E-mail: UCHIDA.Kenichi@nims.go.jp

**Abstract:**

Spin-mediated charge-to-heat current conversion phenomena, i.e., the anomalous Ettingshausen effect (AEE) and the anisotropic magneto-Peltier effect (AMPE), have been investigated in various ferromagnetic Ni-Fe, Ni-Pt, Ni-Pd, and Fe-Pt binary alloys at room temperature. When a charge current is applied to a ferromagnetic conductor, the AMPE modulates the Peltier coefficient depending on the angle between the directions of the charge current and magnetization, while the AEE generates a heat current in the direction perpendicular to both the charge current and magnetization. We observed the strong material dependence of the thermoelectric conversion coefficients and figures of merit of these phenomena. Among the ferromagnetic alloys used in this study, $Ni_{95}Pt_5$ exhibits the largest AMPE of which the anisotropy of the Peltier coefficient is ~12%. In contrast, the magnitude of the AEE signals is moderate in $Ni_{95}Pt_5$ but largest in $Ni_{75}Pt_{25}$ and $Ni_{50}Fe_{50}$. We discuss these behaviors by exploring the relations between these charge-to-heat current conversion phenomena and other transport as well as magnetic properties. This systematic study will provide a clue for clarifying the mechanisms of the AMPE and AEE and for enhancing the thermoelectric conversion efficiency of these phenomena.

## I. INTRODUCTION

The thermoelectric effect refers to the conversion of electricity into heat and vice versa. The conversion of a heat (charge) current into a charge (heat) current in a conductor is called the Seebeck (Peltier) effect [1,2] where the heat and charge currents flow parallel to each other. Materials science research for the last several decades has realized a significant enhancement in the thermoelectric conversion efficiency of the Seebeck and Peltier effects [3-5]. However, to realize the thermoelectric power generation (temperature control) via the Seebeck (Peltier) effect, the junctions of two materials with different Seebeck (Peltier) coefficients are indispensable. This junction structure makes a thermoelectric module complicated, which limits the application of the Seebeck and Peltier effects. Therefore, the investigations on other thermoelectric conversion effects are also necessary for realizing versatile thermal energy harvesting and thermal management applications.

In magnetic materials, in addition to the conventional thermoelectric effects, a variety of magneto-thermoelectric and thermo-spin phenomena appear owing to the spin degree of freedom, which have been intensively investigated in the field of spin caloritronics [6,7]. While early studies on spin caloritronics mainly focused on the transverse heat-to-charge current conversion phenomena, such as the spin Seebeck effect [8-11] and the anomalous Nernst effect (ANE) [12-16], various charge-to-heat current conversion phenomena have recently been observed with developments of thermal measurement techniques. The representative charge-to-heat current conversion phenomena in magnetic conductors are the anisotropic magneto-Peltier effect (AMPE) [17-20] and the anomalous Ettingshausen effect (AEE) [18,20-29]. These phenomena are promising to realize novel and versatile thermal management technologies for spintronic devices because of the different symmetries in thermoelectric conversion and the junction-free structure, as shown below.

The AMPE is the phenomenon that the Peltier coefficient in a ferromagnetic conductor depends on the angle $\theta$ between the directions of the charge current $\mathbf{J}_c$ and magnetization $\mathbf{M}$. The $\theta$ dependence of the Peltier coefficient $\Pi$ in an isotropic ferromagnet is expressed by

$$\Pi(\theta) = \Pi_\perp + (\Pi_\parallel - \Pi_\perp)\cos^2\theta, \qquad (1)$$

where $\Pi_\parallel$ and $\Pi_\perp$ are the Peltier coefficients for the configurations where $\mathbf{J}_c$ is parallel and perpendicular to $\mathbf{M}$, respectively [18]. The cooling and heating due to the AMPE are generated between the areas with different $\theta$ values; non-uniform magnetization distribution along the charge current path in a ferromagnet leads to a temperature change even in the absence of the junctions of two materials, which is different from the situation for the conventional Peltier effect. The existence of the AMPE has long been expected since its reciprocal phenomenon, the anisotropic magneto-Seebeck effect (AMSE) [30-38], was observed and the signature of the AMPE/AMSE appears in non-local spin-transport experiments [17]. In this stream, the direct observation of the AMPE was reported in 2018 [18], where the temperature modulation induced by the AMPE was observed in polycrystalline Ni slabs without any junction structures [Fig. 1(a)] by means of the active infrared emission microscopy called the lock-in thermography (LIT) [39-41]. The AMPE was observed to appear not only in Ni but also in several Ni-based alloys, $Ni_{45}Fe_{55}$, $Ni_{95}Pt_5$, and $Ni_{95}Pd_5$. Here, the magnitude of the AMPE in $Ni_{95}Pt_5$ is several times larger than that in Ni, which is probably owing to the strong spin-orbit interaction in Pt. In contrast, it was also demonstrated that the AMPE in Fe is much smaller than that in Ni. The strong material dependence in the AMPE indicates the important role of the electronic structure of host ferromagnetic metals. A part of these behaviors can be explained by the intrinsic mechanism discussed in Ref. [42].

The AEE refers to the generation of a transverse heat current from a charge current applied to a ferromagnetic conductor and has the following symmetry [25]:

$$\mathbf{j}_{q,\text{AEE}} = \Pi_{\text{AEE}}(\mathbf{j}_c \times \mathbf{m}), \qquad (2)$$

where $\mathbf{j}_{q,\text{AEE}}$, $\Pi_{\text{AEE}}$, $\mathbf{m}$, and $\mathbf{j}_c$ denote the heat current density generated by the AEE, anomalous Ettingshausen coefficient, unit vector of $\mathbf{M}$, and charge current density, respectively. Equation (2) means that a heat current is generated through the AEE in the direction perpendicular to both the charge current and magnetization [Fig. 1(b)]. Although the AEE has also been known for a long time [21-23], its observation in thin films has recently been realized [20,24-28]. Since the AEE enables magnetic control of the heat-current direction, it can be used as a versatile temperature controller if ferromagnets with large $\Pi_{\text{AEE}}$ are found.

In this work, we report the observation of the AMPE and the AEE in various ferromagnetic materials including Ni-Fe, Ni-Pt, Ni-Pd, and Fe-Pt binary alloys at room temperature. To confirm and quantify the behaviors obtained in the AMPE measurements, we also measured the AMSE in these ferromagnets as well as the electrical and thermal conductivities. These measurements allow us to estimate not only the thermoelectric conversion coefficients but also the power factor and figure of merit of these phenomena. Furthermore, the obtained material dependences of the AMPE and the AEE are compared with those of the anisotropic magnetoresistance (AMR) and the anomalous Hall effect (AHE), respectively. This comprehensive study will provide a guideline for clarifying the mechanisms of the AMPE and the AEE and for enhancing thermoelectric conversion efficiency of these phenomena.

This paper is organized as follows. In Sec. II, we explain the details of the experimental procedure and configuration for the measurements of the AMPE and the AEE using the LIT method. In Sec. III, we report the observation of the AMPE/AMSE and the AEE in various ferromagnets and the quantitative estimation of their thermoelectric conversion properties. In Sec. III, we also compare the material dependence of the AMPE/AMSE (AEE) with that of AMR (AHE). The last Sec. IV is devoted to the conclusion of the present study.

## II. SAMPLE PREPARATION AND EXPERIMENTAL PROCEDURE

In this study, we used the polycrystalline Ni and Fe slabs, commercially available from the Nilaco Corporation, Japan, and the polycrystalline $Ni_{95}Fe_5$, $Ni_{75}Fe_{25}$, $Ni_{50}Fe_{50}$, $Ni_{25}Fe_{75}$, $Ni_{95}Pd_5$, $Ni_{70}Pd_{30}$, $Ni_{95}Pt_5$, $Ni_{75}Pt_{25}$, and $Fe_{50}Pt_{50}$ binary alloy slabs prepared by a melting method with rapid cooling, available from Kojundo Chemical Laboratory Co.,Ltd., Japan. From the structural characterization using the x-ray diffraction, it was confirmed that all the alloys are disordered. For the measurements of the AMPE and the AEE, the slabs were cut into a U-shaped structure using fully automatic wire electric discharge machining, where the cutting accuracy is ~3 μm. The length along the $x$ direction, total width along the $z$ direction, width of the leg parts along the $z$ direction, and thickness of the U-shaped slabs along the $y$ direction are 12.0 mm, 1.0 mm, 0.2 mm, and 0.5 mm, respectively [Fig. 2(d)]; the design and dimensions of the U-shaped samples are the same as those used for the previous study on the AMPE [18]. The measurements of the AMSE, electrical conductivity, AMR, and AHE were performed by using the bar-shaped slabs with a size of 12.0 × 1.0 × 0.5 $mm^3$. Both the U-shaped and bar-shaped samples were obtained from the same ingot for each composition.

We measured the temperature modulation induced by the AMPE and the AEE by means of the LIT method [18-20,24-29,39-41]. In the LIT measurements, while applying a square-wave-modulated AC charge current with the frequency $f$ (1-25 Hz) to a sample, the temperature change oscillating with the same frequency as the charge current is extracted. Thermal images obtained using an infrared camera are converted into the lock-in amplitude $A$ and phase $\phi$ images by Fourier analysis, where the conversion procedures from the infrared intensity into the temperature information are detailed in the previous work [18]. The $A$ image shows the spatial distribution of the magnitude of the temperature change in linear response to the oscillating charge current, while the $\phi$ image shows the spatial distribution of the sign of the temperature change in addition to the time delay due to thermal diffusion. In general, the $A$ and $\phi$ images measured at low $f$ values show the temperature distribution at nearly steady states, while those at high $f$ values show the distribution at transient states, where the temperature broadening due to thermal diffusion is suppressed. The LIT measurements at high $f$ values are necessary to determine positions where heating/cooling signals are generated. During the LIT measurements, we applied a magnetic field with the magnitude $|H|$ = 12 kOe along the +$z$ or –$z$ direction, where $|H|$ = 12 kOe is large enough to saturate the magnetization of the U-shaped samples except for Fe (see Appendix A and note that even the magnetization of Fe is almost saturated at $|H|$ = 12 kOe). Here, the magnetic field is uniform over the viewing area, because the size of the magnetic pole pieces of our electromagnet is much bigger than the regions of interest. We also found that the tilt angles of the samples and the electromagnet are smaller than 1° for all the LIT measurements, confirming that the variations of the AMPE and AEE signals due to the deviation of the angle between the magnetic field and charge current is negligibly small. Since the AMPE (AEE) exhibits the even (odd) dependence on the $H$ sign [Eqs. (1) and (2)], one can extract the AMPE and the AEE contributions from the thermal images measured at the positive and negative fields. The AMPE signals can be separated from the background due to the $H$-independent Peltier effect by subtracting the thermal images measured at $H$ = 0 kOe from those at $|H|$ = 12 kOe. To improve infrared emissivity, the surface of the samples was uniformly coated with insulating black ink. The samples used for the LIT measurements were fixed on a plastic plate with low thermal conductivity to reduce the heat loss from the samples to surroundings as much as possible. All the measurements were carried out at room temperature and atmospheric pressure.

We also investigated the AMSE of the ferromagnetic metal slabs at room temperature. A temperature gradient $\nabla T$ was applied between the ends of the bar-shaped slab by attaching one end of the slab to a chip heater and the other end to a heat bath. We measured the voltage difference $\Delta V$ between the ends of the slab with applying $H$ along the temperature gradient; from the $H$ dependence of the thermoelectric voltage, the AMSE contribution can be extracted. Here, the Seebeck coefficient of the slabs at $H$ = 0 kOe was calibrated by using the Seebeck Coefficient/Electric Resistance Measurement System (ZEM-3, ADVANCE RIKO, Inc.), where the electrical conductivity was also measured simultaneously.

In addition to the aforementioned magneto-thermoelectric effects, we measured the static magnetic properties and transport coefficients of the ferromagnetic metal slabs systematically. For estimating the AHE (AMR), the transverse (longitudinal) resistivity of the bar-shaped slabs was measured by a four-probe method; the voltage difference along the 0.5 mm (12.0 mm) direction of the slabs was measured with sweeping $H$ from −10 kOe (−6 kOe) to 10 kOe (6 kOe) in the 1.0 mm (12.0 mm) direction.

## III. RESULTS AND DISCUSSION
### A. Separation of AMPE and AEE

Figure 2(a) shows the $A$ and $\phi$ images for the $Ni_{95}Pd_5$ slab at $J_c$ = 1.0 A and $f$ = 25 Hz. A clear temperature change induced by a charge current appears at $H$ = 12 kOe. This temperature change is irrelevant to the conventional Peltier effect because it disappears in the absence of a magnetic field, where the remanent magnetization is negligibly small (Appendix A). These results indicate that the current-induced temperature change at finite magnetic fields originates from the AMPE and/or AEE of which the field dependence follows the magnetization curve, as shown in the previous work [18]. When the magnetic field of $H$ = −12 kOe was applied, a different pattern was observed due to the difference in the symmetry between the AMPE and the AEE; the AMPE (AEE) shows the even (odd) dependence on the direction of the magnetization [18,20].

To separate these two contributions, we extracted the $H$-odd ($H$-even) component of the LIT images, which is obtained by subtracting (adding) the raw LIT images at $H$ = −12 kOe from (to) those at $H$ = 12 kOe and dividing the subtracted (added) images by 2. As shown in Fig. 2(b), the current-induced temperature change with the $H$-odd dependence appears in the leg parts of the U-shaped slab, where the charge current direction is perpendicular to the magnetization. The phase difference between the left and right legs was observed to be ~180°, indicating that the sign of the temperature change on the surface of the left leg is opposite to that of the right leg. These features are consistent with the symmetry of the AEE [Eq. (2)] [24,25]. In contrast, the $H$-even component is composed of not only the AMPE signal but also the background due to the $H$-independent Peltier effect [20]. Therefore, we obtained the pure AMPE contribution by subtracting the raw LIT images at $H$ = 0 kOe from the $H$-even component. Although the $H$-independent background in the $Ni_{95}Pd_5$ slab is negligibly small [compare Figs. 2(c) with 2(e)], this analysis allows us to extract the pure AMPE contribution even when the finite background signal appears due to the crystal orientation dependence of the Peltier coefficient. Figure 2(e) shows that the pure AMPE signal in the $Ni_{95}Pd_5$ slab appears around the corners of the U-shaped slab and the phase difference between the left and right corners is ~180°, consistent with the previous work [18]. We extracted the pure AMPE and the AEE contributions in the other samples in the same way. Hereafter, the amplitude and phase images showing the pure AMPE (AEE) contribution are denoted by $A_{AMPE(AEE)}$ and $\phi_{AMPE(AEE)}$, respectively.

### B. AMPE in various ferromagnetic metals

Figure 3(a) shows the $A_{AMPE}$ and $\phi_{AMPE}$ images in various ferromagnetic metal slabs at $J_c$ = 1.0 A, $|H|$ = 12 kOe, and $f$ = 25 Hz. We observed clear AMPE signals around the corners of the U-shaped structure in Ni, Ni-Pt, Ni-Pd, and Ni-Fe alloys. Interestingly, the AMPE signal is negligibly small not only in Fe but also in $Fe_{50}Pt_{50}$ despite its strong spin-orbit interaction, which is qualitatively consistent with the results of the first-principles calculations [42]. These LIT-based measurements confirm the versatility of the AMPE in Ni-based alloys and emphasize the strong composition dependence of this phenomenon.

Figures 3(b) and 3(c) show the $f$ dependence of $A_{AMPE}$ at the left corner of the U-shaped ferromagnets. The amplitude of the temperature modulation monotonically increases with decreasing $f$ as observed in the previous work [18,20]. We found that the $f$ dependence of $A_{AMPE}$ depends on the composition [the inset to Fig. 3(c)]; the $f$ dependence for $Ni_{50}Fe_{50}$, $Ni_{25}Fe_{75}$, and $Ni_{75}Pt_{25}$ are greater than that for the other ferromagnets. This is caused by the difference in the thermal diffusion length or the thermal conductivity $\kappa$. In fact, due to alloying, the magnitude of $\kappa$ for $Ni_{50}Fe_{50}$, $Ni_{25}Fe_{75}$, and $Ni_{75}Pt_{25}$ is several times smaller than that for Ni (Appendix B). Although the temperature modulation in the steady-state condition is necessary to quantitatively estimate the magnitude of the AMPE coefficient, it is difficult to quantify the steady-state AMPE signals from the LIT measurements alone because the exact distribution of the charge current around the corners is not clear in our samples and heat loss to surroundings increases with decreasing $f$.

To quantify the anisotropy of the Peltier coefficients in the ferromagnets, we measured the AMSE. The Onsager reciprocal relation between the AMPE and the AMSE indicates $\Delta S = S_\parallel - S_\perp = (\Pi_\parallel - \Pi_\perp)/T$, where $S_\parallel$ ($S_\perp$) is the Seebeck coefficient for the $\mathbf{M} \parallel \nabla T$ ($\mathbf{M} \perp \nabla T$) configuration [18] and $T$ is the temperature. In a similar manner to the AMR in an isotropic ferromagnetic metal [43], we estimated the magneto-Seebeck coefficient as $\Delta S = 3(S_{|H| = 2\,kOe} - S_{xx})/2$. Here, $S_{|H| = 2\,kOe}$ denotes the Seebeck coefficient of the ferromagnets at $|H|$ = 2 kOe (note that the $\mathbf{M}$ direction of the bar-shaped ferromagnets aligns along the field direction at $|H|$ = 2 kOe in our AMSE setup). $S_{xx}$ denotes the Seebeck coefficient in the absence of the magnetic field, of which the material dependence is shown in Appendix B. We found that the thermoelectric voltage in all the ferromagnets shows the $H$-even dependence when the magnetic field is applied along the $\nabla T$ direction, indicating that the $H$ dependence of $S$ is dominated by the AMSE and the contribution of the ANE is negligibly small due to the collinear orientation of **M** and $\nabla T$ [the inset to Fig. 4(b)]. As shown in Fig. 4(a), the sign of $\Delta S$ is positive for the Ni-based alloys and its magnitude strongly depends on the composition. The material dependence of $\Delta S$ is similar to that of the AMPE signals measured in the LIT experiments.

Since the AMSE signals observed here reflect the steady-state transport properties, they can be directly used for estimating the thermoelectric conversion performance. Figure 4(b) shows the AMSE ratio, $\Delta S/|S_{xx}|$, for the ferromagnets. The AMSE ratio for Ni is estimated to be 3%, which is almost the same as the value reported in Ref. [18]. The $\Delta S/|S_{xx}|$ values for most of the ferromagnets are comparable to or smaller than the value for Ni. Importantly, $Ni_{95}Pt_5$ exhibits the prominently large AMSE ratio: $\Delta S/|S_{xx}| = \sim 12\%$, which is about 4 times larger than that for Ni. We also estimated the dimensionless figure of merit for the AMSE:

$$Z_{\text{AMSE}}T = \frac{\Delta S^2 \sigma_{xx}}{\kappa} T, \qquad (3)$$

where $\sigma_{xx}$ denotes the longitudinal electrical conductivity. As a natural consequence of the relation $Z_{\text{AMSE}}T \propto \Delta S^2$ and the strong material dependence of $\Delta S$, the material dependence of $Z_{\text{AMSE}}T$ is further pronounced [Fig. 4(b)]. The maximum $Z_{\text{AMSE}}T$ value is obtained for $Ni_{95}Pt_5$; the estimated value at $T = 300$ K is $1.0 \times 10^{-4}$. This value is one order of magnitude greater than that of Ni but two orders of magnitude smaller than the figure of merit of the conventional Seebeck effect (Appendix B).

Figure 5 shows the dependence of $\Delta S$ on the saturation magnetization $M_s$ of the ferromagnets. We found that, in the Ni-Fe alloys, the magnitude of $\Delta S$ has a negative correlation with $M_s$ (see the pink data points in Fig. 5). In contrast, the AMSE in the Ni-Pt and Ni-Pd alloys clearly deviates from the $M_s$-$\Delta S$ correlation for the Ni-Fe alloys; the $\Delta S$ value for $Ni_{95}Pt_5$ and $Ni_{95}Pd_5$ is larger than that for Ni although $M_s$ of these materials is comparable to each other, emphasizing the essential role of the spin-orbit interaction in the AMSE (Fig. 5). However, excess Pt and Pd contents decrease the magnitude of $\Delta S$ (Fig. 4). These results indicate that doping of heavy elements with strong spin-orbit interaction to ferromagnetic materials is useful to improve the thermoelectric conversion efficiency of the AMSE but its material dependence for such alloys is not simple, which cannot be predicted by the static magnetic property.

To provide a clue for understanding the origin of the observed material dependence, we examined the relation between the AMSE and the AMR in the ferromagnets. Here, let us remind you that the Seebeck coefficient in a simple metal is given by the Mott relation [42]:

$$S_{xx} = -\frac{\pi^2 k_B^2 T}{3e} \frac{1}{\sigma_{xx}(\varepsilon_F)} \left[ \frac{\partial \sigma_{xx}(\varepsilon)}{\partial \varepsilon} \right]_{\varepsilon=\varepsilon_F}, \qquad (4)$$

where $k_B$ is the Boltzmann constant, $e$ ($> 0$) the elementary charge, $\sigma_{xx}(\varepsilon)$ the energy ($\varepsilon$)-dependent electrical conductivity, and $\varepsilon_F$ the Fermi energy. Based on Eq. (4), the AMSE ratio can be described as

$$\frac{\Delta S}{S_{xx}} \approx \left(1 - \frac{\sigma_\parallel}{\sigma_\perp}\right) - \left\{1 - \frac{[\partial \sigma_\parallel(\varepsilon)/\partial \varepsilon]_{\varepsilon=\varepsilon_F}}{[\partial \sigma_\perp(\varepsilon)/\partial \varepsilon]_{\varepsilon=\varepsilon_F}}\right\} = \left(1 - \frac{\sigma_\parallel}{\sigma_\perp}\right) - \left(1 - \frac{\sigma_\parallel S_\parallel}{\sigma_\perp S_\perp}\right), \qquad (5)$$

when $\Delta S \ll S_\parallel, S_\perp$ and $|\sigma_\parallel - \sigma_\perp| \ll \sigma_\parallel, \sigma_\perp$ with $\sigma_\parallel$ ($\sigma_\perp$) being the electrical conductivity for the **M** $\parallel \nabla T$ (**M** $\perp \nabla T$) configuration. The first term of Eq. (5), $1 - \sigma_\parallel/\sigma_\perp$, represents the contribution of the AMR in the AMSE, while the second term, $1 - \sigma_\parallel S_\parallel/(\sigma_\perp S_\perp)$, represents the contribution coming from the anisotropy of the $\varepsilon$ derivative of the electrical conductivity at $\varepsilon = \varepsilon_F$. Here, both the first and second terms become zero when the transport properties are isotropic. To clarify the contribution of each term in Eq. (5), we plot the $1 - \sigma_\parallel/\sigma_\perp$ and $1 - \sigma_\parallel S_\parallel/(\sigma_\perp S_\perp)$ values for the ferromagnets in Fig. 6. The $\sigma_\parallel/\sigma_\perp$ values for all the ferromagnets are smaller than 3% because of the small AMR (Fig. 6 and Appendix B). Importantly, only $Ni_{95}Pt_5$ shows the remarkably large $1 - \sigma_\parallel S_\parallel/(\sigma_\perp S_\perp)$ value of ~13% (Fig. 6). This result indicates that the large AMSE in $Ni_{95}Pt_5$ is attributed to the anisotropy of the $\varepsilon$ derivative of the electrical conductivity at $\varepsilon = \varepsilon_F$. In fact, the first-principles study on the AMPE/AMSE reported previously [42], which is based on Eq. (4) combined with the spin-orbit interaction, successfully explains the experimental behaviors for Ni and Fe, suggesting the important contribution coming from the $\varepsilon$ derivative of the electrical conductivity.

Although we performed the systematic measurements on the AMPE/AMSE and related transport properties, it is still difficult to obtain clear strategies to enhance the anisotropy of the Pelteir/Seebeck coefficient and predict its material dependence. Nevertheless, our experiments clearly show the essential role of the anisotropy of the $\varepsilon$ derivative of the electrical conductivity at $\varepsilon = \varepsilon_F$ in the AMPE/AMSE. We also found that doping of heavy elements with strong spin-orbit interaction to ferromagnetic metals is effective but the doping amount should not be too much, which was confirmed by the fact that the AMPE/AMSE in our $Ni_{95}Pt_5$ and $Ni_{95}Pd_5$ ($Ni_{75}Pt_{25}$ and $Ni_{70}Pd_{30}$) samples is greater (smaller) than that in Ni (Fig. 4). These results are useful for designing materials that show large anisotropy of the Peltier/Seebeck coefficient. Furthermore, the aforementioned first-principles study [42] predicts that $L1_2$-ordered $Ni_{75}Pt_{25}$ and $L1_0$-ordered $Ni_{50}Pt_{50}$ show large AMPE/AMSE, while we focus on the disordered alloys in this work. We thus anticipate that the AMPE/AMSE in ordered alloys will enhance the thermoelectric output and the investigation of them will deepen understanding of the material dependence of the thermoelectric conversion properties.

### C. AEE in various ferromagnetic metals

Figure 7(a) shows the $A_{AEE}$ and $\phi_{AEE}$ images in various ferromagnetic metal slabs at $J_c = 1.0$ A, $|H| = 12$ kOe, and $f = 25$ Hz. We observed clear AEE signals in all the samples, of which the spatial distribution is consistent with Eq. (2). The magnitude of the AEE signals strongly depends on the material species [see the $A_{AEE}$ images in Fig. 7(a)]. The sign of the AEE signals for all the binary alloys is the same as that for Ni, while only Fe exhibit the AEE signal with the opposite sign [see the $\phi_{AEE}$ images in Fig. 7(a)].

Figures 7(b) and 7(c) show the $f$ dependence of the AEE signals on the left leg of the ferromagnets. In a similar manner to the AMPE signals, the amplitude of the AEE-induced temperature change monotonically increases with decreasing $f$ (note that the AEE signals in in-plane magnetized thin films are frequency-independent because of the size effect [20]). Although the $f$ dependence of the AEE signals is rather weak, we have to estimate their magnitude in the steady state to obtain the AEE coefficient. Therefore, we combine the experimental results obtained from the LIT measurements with the calculated $f$ dependence of the AEE signals based on the one-dimensional heat diffusion equation in the frequency domain, which allows us to estimate the magnitude of the AEE-induced temperature change at $f = 0$ Hz. Here, as a model system for the calculation, we consider a ferromagnetic metal with the length $L = 0.5$ mm along the direction of the AEE-induced heat current. The total heat flux at the ends of the ferromagnetic metal is set to be zero; this is the boundary condition for solving the heat diffusion equation. The temperature modulation $A_\mathrm{AEE}$ at the end of the ferromagnetic metal is given by

$$A_\mathrm{AEE} e^{-i\phi} = \sqrt{\frac{i}{2\pi f C \rho \kappa}} \frac{\cos\left(-L\sqrt{2\pi f C \rho / i\kappa}\right) - 1}{\sin\left(-L\sqrt{2\pi f C \rho / i\kappa}\right)} \tilde{j}_\mathrm{q,AEE}, \tag{6}$$

where $C$ is the specific heat capacity, $\rho$ the density, and $\tilde{j}_\mathrm{q,AEE}$ the first-harmonic sinusoidal amplitude of the heat current density generated by the AEE [29]. By substituting the experimental values of $\kappa$, $C$, and $\rho$ into Eq. (6), we can calculate the $f$ dependence of the AEE signal, which well reproduces the experimental behavior in Figs. 7(b) and 7(c). From the experimental and calculation results, we then estimated the AEE coefficient by using the following equation based on Eq. (2) [24,25,29]:

$$\Pi_\mathrm{AEE} = \frac{\pi \kappa \Delta T}{4 j_\mathrm{c} L}, \tag{7}$$

where $\Delta T = 2A_\mathrm{AEE}\cos\phi_\mathrm{AEE}$ and $(4/\pi)j_\mathrm{c}$ is the sinusoidal amplitude of the charge current density applied to the ferromagnet.

In Fig. 8(a), we show $\Pi_\mathrm{AEE}$ and the corresponding ANE coefficient $S_\mathrm{ANE}$ ($= \Pi_\mathrm{AEE}/T$) at $T = 300$ K in various ferromagnetic metals. The AEE coefficient of the Ni-Fe alloys with Ni-rich composition is almost independent of the Fe concentration while the AEE is significantly enhanced in $Ni_{50}Fe_{50}$. The magnitude of $\Pi_\mathrm{AEE}$ for $Ni_{25}Fe_{75}$ is smaller than that for $Ni_{50}Fe_{50}$ but still much larger than that for the Ni-rich alloys. Significantly, despite the absence of heavy elements, the magnitude of $\Pi_\mathrm{AEE}$ for $Ni_{50}Fe_{50}$ is larger than that for the Ni-Pt, Ni-Pd, and Fe-Pt alloys that are expected to possess the strong spin-orbit interaction. In the Ni-Pt and Ni-Pd alloys, the AEE coefficient increases as the concentration of the heavy elements increases, which is an opposite trend to the material dependence of the AMPE/AMSE. Our results show that, although both the AEE and the AMPE/AMSE originate from the spin-polarized electron transport affected by the spin-orbit interaction, their behavior is completely different from each other. As shown in Fig. 9(a), $\Pi_\mathrm{AEE}$ of the ferromagnets has no correlation with $\Delta S$, which represents the performance of the AMPE/AMSE. We also note that $\Pi_\mathrm{AEE}$ has no correlation with $M_\mathrm{s}$ [Fig. 9(b)]; the scaling between the magnetization and transverse thermoelectric effects discussed in the previous study [14] is inapplicable even to the simple binary alloys.

Figure 8(b) shows the dimensionless figure of merit of the AEE in the ferromagnetic metals, which is defined as the following equation [29,44]:

$$Z_\mathrm{AEE} T = \frac{\Pi_\mathrm{AEE}^2 \sigma_{xx}}{\kappa} \frac{1}{T}. \tag{8}$$

Among the ferromagnets used in this study, $Ni_{50}Fe_{50}$ exhibits the largest figure of merit of $Z_\mathrm{AEE}T = 2.8 \times 10^{-5}$ at $T = 300$ K, which is two orders of magnitude greater than that of Ni but comparable to that of $Ni_{75}Pt_{25}$. Here, we also note that even the maximum $Z_\mathrm{AEE}T$ value for the ferromagnetic binary alloys used in this study is one order of magnitude smaller than the record-high values [15,29].

Now we move to discuss the origin of the strong material dependence of the AEE. The AEE coefficient of a ferromagnet can be divided into the following two terms [16,29]:

$$\Pi_\mathrm{AEE} = \left(\rho_{xx}\alpha_{xy} + \rho_{xy}\alpha_{xx}\right)T \equiv \Pi_\mathrm{I} + \Pi_\mathrm{II}, \tag{9}$$

where $\rho_{xx} = 1/\sigma_{xx}$ ($\rho_{xy} = -\sigma_{xy}/\sigma_{xx}^2$) is the diagonal (off-diagonal) component of the electrical resistivity tensor, $\alpha_{xx}$ ($\alpha_{xy}$) is the diagonal (off-diagonal) component of the thermoelectric conductivity tensor, $\Pi_{\mathrm{I}} = \rho_{xx}\alpha_{xy}T$, and $\Pi_{\mathrm{II}} = \rho_{xy}\alpha_{xx}T$. The $\Pi_{\mathrm{I}}$ term is the intrinsic part of the AEE originating from the transverse thermoelectric conductivity $\alpha_{xy}$, which is given by [12,15]

$$\alpha_{xy} = -\frac{\pi^2 k_{\mathrm{B}}^2 T}{3e}\left[\frac{\partial \sigma_{xy}(\varepsilon)}{\partial \varepsilon}\right]_{\varepsilon=\varepsilon_{\mathrm{F}}}. \tag{10}$$

In contrast, the $\Pi_{\mathrm{II}}$ term is expressed as $\Pi_{\mathrm{II}} = S_{xx}T\tan\theta_{\mathrm{AHE}}$ with $S_{xx} = \rho_{xx}\alpha_{xx}$ and $\theta_{\mathrm{AHE}} = \rho_{xy}/\rho_{xx}$ respectively being the Seebeck coefficient and the anomalous Hall angle. We estimated $\Pi_{\mathrm{I}}$ and $\Pi_{\mathrm{II}}$ of the various ferromagnetic materials following the procedures in Ref. [29] and summarized the results in Fig. 10. Interestingly, although the thermoelectric performance of the AEE of $Ni_{50}Fe_{50}$ and $Ni_{75}Pt_{25}$ is comparable to each other, the origin is different. The AEE of $Ni_{75}Pt_{25}$ is dominated by the intrinsic $\Pi_{\mathrm{I}}$ term; the large $\Pi_{\mathrm{I}}$ value comes from the combination of the considerably large $\alpha_{xy}$ and low $\sigma_{xx}$ values [Fig. 10(b) and Appendix B]. However, the $\alpha_{xy}$ value of $Ni_{75}Pt_{25}$ is comparable to that of pure Ni and other Ni-Pt and Ni-Pd alloys, indicating that the doping of heavy elements is not effective for improving $\alpha_{xy}$ in the case of the simple Ni-based alloys. In contrast, in $Ni_{50}Fe_{50}$, both the $\Pi_{\mathrm{I}}$ and $\Pi_{\mathrm{II}}$ terms provide significant contributions; although $\alpha_{xy}$ of $Ni_{50}Fe_{50}$ is smaller than that of Ni, Ni-Pt, and Ni-Pd alloys, the large $S_{xx}$ value makes the $\Pi_{\mathrm{II}}$ term important in $Ni_{50}Fe_{50}$. These results suggest that the improvement of $\alpha_{xy}$ is not only the way to enhance the AEE, and that the effective use of the Seebeck effect and the AHE in ferromagnetic metals is also important [16].

Finally, we discuss the correlation between the transport coefficients determining the AEE. Since $\alpha_{xy}$ can be dependent on $\sigma_{xy}$ through Eq. (10), we plot the $\alpha_{xy}$ values for the ferromagnets as a function of $\sigma_{xy}$ [Fig. 11(a)]. We found a weak but clear correlation between these parameters; the $\alpha_{xy}$ values for the ferromagnetic metals with positive $\sigma_{xy}$ are larger than those with negative $\sigma_{xy}$ and tend to increase with increasing $\sigma_{xy}$. The observed correlation between $\sigma_{xy}$ and $\alpha_{xy}$ forms the basis for establishing the scaling behavior between the transverse electrical and thermoelectric conductivities, while more systematic measurements using various classes of materials in a wide range of temperatures are necessary. We also note that the longitudinal and transverse electrical conductivities in our ferromagnets is consistent with the scaling relation for the AHE [12], as shown in Fig. 11(b) (see also the $\theta_{\mathrm{AHE}}$ values shown in Appendix B).

## IV. CONCLUSION

In summary, we systematically investigated the magneto-thermoelectric effects and related transport properties in various ferromagnetic metals including Ni-Fe, Ni-Pt, Ni-Pd, and Fe-Pt binary alloys at room temperature. Based on the experimental results, we estimated the thermoelectric conversion coefficients as well as the figures of merit of the AMPE/AMSE and the AEE quantitatively. The figures of merit of the AMPE/AMSE and the AEE at room temperature are more than two orders of magnitude smaller than those of the conventional Seebeck effect, indicating that one order of magnitude improvement in the thermoelectric conversion coefficients is at least necessary for thermoelectric applications. We observed clear AMPE/AMSE signals in most of the ferromagnetic metals and found that not only the AMR contribution but also the anisotropy of the energy derivative of the electrical conductivity at the Fermi energy is important for the AMPE/AMSE, which provides a clue for obtaining materials with a large AMPE/AMSE coefficient. By measuring the material dependence of the AEE using the ferromagnetic binary alloys, we found a correlation between the anomalous Hall conductivity and the transverse thermoelectric conductivity. Our experimental results also show that the AEE can be enhanced not only by improving the transverse thermoelectric conductivity but also by utilizing the large Seebeck coefficient and the anomalous Hall conductivity. To clarify the microscopic mechanism of the transport coefficients and to determine an appropriate temperature range for the use of the magneto-thermoelectric effects, the temperature dependence of the AMPE/AMSE and the AEE needs to be investigated, while this work focuses only on the room-temperature measurements. Comprehensive material dependence measurements are also important to further improve the thermoelectric performance of the magneto-thermoelectric effects, where combinatorial materials science will be useful [45,46]. We anticipate that the systematic data set reported here will provide a hint for future studies.

## ACKNOWLEDGMENT

The authors thank K. Masuda and Y. Miura for valuable discussions and M. Isomura and Materials Processing Group, Materials Manufacturing and Engineering Station, National Institute for Materials Science, for technical supports in sample preparation. This work was supported by CREST "Creation of Innovative Core Technologies for Nano-enabled Thermal Management" (JPMJCR17I1) from JST, Japan, Grant-in-Aid for Scientific Research (B) (JP19H02585), Grant-in-Aid for Scientific Research (S) (JP18H05246), and Grant-in-Aid for Early-Career Scientists (JP18K14116) from JSPS KAKENHI, Japan, and the NEC Corporation. A.M. is supported by JSPS through a research fellowship for young scientists (JP18J02115).

## APPENDIX A: MAGNETIZATION CURVE

The magnetization curves of the samples were measured by the vibrating sample magnetometry at room temperature with applying the magnetic field with the magnitude $H$ ranging from $-15$ kOe to $15$ kOe to the U-shaped and bar-shaped ferromagnetic metal slabs, where the field direction is depicted in the inset to Fig. 12. The saturation magnetic field for the U-shaped samples was observed to be larger than that for the bar-shaped samples due to the difference in the shape magnetic anisotropy. The transport measurements shown in the main text were performed in the condition where $H$ is large enough to saturate the magnetization of the samples, except for the U-shaped Fe sample (note that the maximum magnetic field available in our LIT system, $|H| = 12$ kOe, is slightly smaller than the saturation field for the U-shaped Fe slab). However, since the magnetic field dependence of the AMPE and AEE signals follows the magnetization curve [18], no big change in the AMPE and the AEE data appears even when the stronger field is applied to the samples.

## APPENDIX B. ELECTRICAL, THERMAL, AND THERMOELECTRIC TRANSPORT PROPERTIES

In Figs. 13(a)-13(c), we show the $H$-independent Seebeck coefficient $S_{xx}$, longitudinal electrical conductivity $\sigma_{xx}$, and thermal conductivity $\kappa$ of the ferromagnets used in this study at room temperature, respectively. By using the measured transport coefficients, we also estimated the figure of merit for the Seebeck effect, $Z_{SE}T = S_{xx}^2 \sigma_{xx} T/\kappa$, at $T = 300$ K [Fig. 13(d)]. As described in Sec. II, $S_{xx}$ and $\sigma_{xx}$ of the bar-shaped ferromagnetic metal slabs were measured by using the Seebeck Coefficient/Electric Resistance Measurement

System (ZEM-3, ADVANCE RIKO, Inc.). We found that the obtained composition dependence of $S_{xx}$ for the Ni-Fe alloys is consistent with the results shown in Ref. [47]. The thermal conductivity was estimated from the thermal diffusivity and specific heat capacity measured using the laser flash system (LFA 1000, Linseis Messgeräte GmbH). The sample size used for the thermal conductivity measurements is $10.0 \times 10.0 \times 0.5$ mm$^3$.

Figures 14(a) and 14(b) show $\Delta\rho/\rho_{xx}$ and $\theta_{\mathrm{AHE}}$ for the ferromagnets, respectively. The Ni-content dependence of $\Delta\rho/\rho_{xx}$ for the Ni-Fe alloys is quantitatively consistent with the results shown in Ref. [48].

**References**

[1]  T. J. Seebeck, Abh. Königl. Akad. Wiss. Berlin **1822**, 265 (1822).

[2]  J. C. A. Peltier, Ann. Chim. Phys. **56**, 371 (1834).

[3]  B. Poudel, Q. Hao, Y. Ma, Y. C. Lan, A. Minnich, B. Yu, X. A. Yan, D. Z. Wang, A. Muto, D. Vashaee, X. Y. Chen, J. M. Liu, M. S. Dresselhaus, G. Chen, and Z. F. Ren, Science **320**, 634 (2008).

[4]  K. Biswas, J. Q. He, Q. C. Zhang, G. Y. Wang, C. Uher, V. P. Dravid, and M. G. Kanatzidis, Nat. Chem. **3**, 160 (2011).

[5]  K. Biswas, J. Q. He, I. D. Blum, C. I. Wu, T. P. Hogan, D. N. Seidman, V. P. Dravid, and M. G. Kanatzidis, Nature **489**, 414 (2012).

[6]  G. E. W. Bauer, E. Saitoh, and B. J. van Wees, Nat. Mater. **11**, 391 (2012).

[7]  S. R. Boona, R. C. Myers, and J. P. Heremans, Energy Environ. Sci. **7**, 885 (2014).

[8]  K. Uchida, S. Takahashi, K. Harii, J. Ieda, W. Koshibae, K. Ando, S. Maekawa, and E. Saitoh, Nature **455**, 778 (2008).

[9]  K. Uchida, J. Xiao, H. Adachi, J. Ohe, S. Takahashi, J. Ieda, T. Ota, Y. Kajiwara, H. Umezawa, H. Kawai, G. E. W. Bauer, S. Maekawa, and E. Saitoh, Nat. Mater. **9**, 894 (2010).

[10] C. Jaworski, J. Yang, S. Mack, D. Awschalom, J. Heremans, and R. Myers, Nat. Mater. **9**, 898 (2010).

[11] K. Uchida, H. Adachi, T. Ota, H. Nakayama, S. Maekawa, and E. Saitoh, Appl. Phys. Lett. **97**, 172505 (2010).

[12] T. Miyasato, N. Abe, T. Fujii, A. Asamitsu, S. Onoda, Y. Onose, N. Nagaosa, and Y. Tokura, Phys. Rev. Lett. **99**, 086602 (2007).

[13] Y. Sakuraba, K. Hasegawa, M. Mizuguchi, T. Kubota, S. Mizukami, T. Miyazaki, and K. Takanashi, Appl. Phys. Express **6**, 033003 (2013).

[14] M. Ikhlas, T. Tomita, T. Koretsune, M. Suzuki, D. Nishio-Hamane, R. Arita, Y. Otani, and S. Nakatsuji, Nat. Phys. **13**, 1085 (2017).

[15] A. Sakai, Y. P. Mizuta, A. A. Nugroho, R. Sihombing, T. Koretsune, M. Suzuki, N. Takemori, R. Ishii, D. Nishio-Hamane, R. Arita, P. Goswami, and S. Nakatsuji, Nat. Phys. **14**, 1119 (2018).

[16] Y. Sakuraba, K. Hyodo, A. Sakuma, and S. Mitani, Phys. Rev. B (accepted).

[17] K. S. Das, F. K. Dejene, B. J. van Wees, and I. J. Vera-Marun, Phys. Rev. B **94**, 180403(R) (2016).

[18] K. Uchida, S. Daimon, R. Iguchi, and E. Saitoh, Nature **558**, 95 (2018).

[19] R. Das and K. Uchida, Appl. Phys. Lett. **114**, 082401 (2019).

[20] R. Das, R. Iguchi, and K. Uchida, Phys. Rev. Appl. **11**, 034022 (2019).

[21] P. W. Bridgman, Phys. Rev. **24**, 644 (1924).

[22] E. H. Hall, Phys. Rev. **26**, 820 (1925).

[23] E. Butler Jr and E. Pugh, Phys. Rev. **57**, 916 (1940).

[24] T. Seki, R. Iguchi, K. Takanashi, and K. Uchida, Appl. Phys. Lett. **112**, 152403 (2018).

[25] T. Seki, R. Iguchi, K. Takanashi, and K. Uchida, J. Phys. D: Appl. Phys. **51**, 254001 (2018).

[26] R. Iguchi, A. Yagmur, Y. C. Lau, S. Daimon, E. Saitoh, M. Hayashi, and K. Uchida, Phys. Rev. B **98**, 014402 (2018).

[27] T. Seki, A. Miura, K. Uchida, T. Kubota, and K. Takanashi, Appl. Phys. Express **12**, 023006 (2019).

[28] S. Ota, K. Uchida, R. Iguchi, P. Van Thach, H. Awano, and D. Chiba, Sci. Rep. **9**, 13197 (2019).

[29] A. Miura, H. Sepehri-Amin, K. Masuda, H. Tsuchiura, Y. Miura, R. Iguchi, Y. Sakuraba, J. Shiomi, K. Hono, and K. Uchida, Appl. Phys. Lett. **115**, 222403 (2019).

[30] J. P. Jan, in *Solid State Physics* (Elsevier, New York, 1957), pp. 1-96.

[31] G. Grannemann and L. Berger, Phys. Rev. B **13**, 2072 (1976).

[32] J.-E. Wegrowe, Q. A. Nguyen, M. Al-Barki, J. F. Dayen, T. L. Wade, and H. J. Drouhin, Phys. Rev. B **73**, 134422 (2006).

[33] A. D. Avery, R. Sultan, D. Bassett, D. Wei, and B. L. Zink, Phys. Rev. B **83**, 100401 (2011).

[34] R. Mitdank, M. Handwerg, C. Steinweg, W. Töllner, M. Daub, K. Nielsch, and S. Fischer, J. Appl. Phys. **111**, 104320 (2012).

[35] A. D. Avery, M. R. Pufall, and B. L. Zink, Phys. Rev. Lett. **109**, 196602 (2012).

[36] A. D. Avery, M. R. Pufall, and B. L. Zink, Phys. Rev. B **86**, 184408 (2012).

[37] M. Schmid, S. Srichandan, D. Meier, T. Kuschel, J.-M. Schmalhorst, M. Vogel, G. Reiss, C. Strunk, and C. H. Back, Phys. Rev. Lett. **111**, 187201 (2013).

[38] O. Reimer, D. Meier, M. Bovender, L. Helmich, J.-O. Dreessen, J. Krieft, A. S. Shestakov, C. H. Back, J.-M. Schmalhorst, A. Hütten, G. Reiss, and T. Kuschel, Sci. Rep. **7**, 40586 (2017).

[39] O. Breitenstein, W. Warta, and M. Langenkamp, *Lock-in thermography: Basics and Use for Evaluating Electronic Devices and Materials Introduction* (Springer, Berlin/Heidelberg, Berlin/Heidelberg, 2010).

[40] O. Wid, J. Bauer, A. Muller, O. Breitenstein, S. S. P. Parkin, and G. Schmidt, Sci. Rep. **6**, 28233 (2016).

[41] S. Daimon, R. Iguchi, T. Hioki, E. Saitoh, and K. Uchida, Nat. Commun. **7**, 13754 (2016).

[42] K. Masuda, K. Uchida, R. Iguchi, and Y. Miura, Phys. Rev. B **99**, 104406 (2019).

[43] T. McGuire and R. Potter, IEEE T. Magn. **11**, 1018 (1975).

[44] K. Uchida, H. Adachi, T. Kikkawa, A. Kirihara, M. Ishida, S. Yorozu, S. Maekawa, and E. Saitoh, Proc. IEEE **104**, 1946 (2016).

[45] I. Takeuchi, O. O. Famodu, J. C. Read, M. A. Aronova, K.-S. Chang, C. Craciunescu, S. E. Lofland, M. Wuttig, F. C. Wellstood, L. Knauss, and A. Orozco, Nat. Mater. **2**, 180 (2003).

[46] K. Uchida, M. Sasaki, Y. Sakuraba, R. Iguchi, S. Daimon, E. Saitoh, and M. Goto, Sci. Rep. **8**, 16067 (2018).

[47] T. Maeda and T. Somura, J. Phys. Soc. Jpn. **44**, 148 (1978).

[48] R. M. Bozorth, Phys. Rev. **70**, 923 (1946).

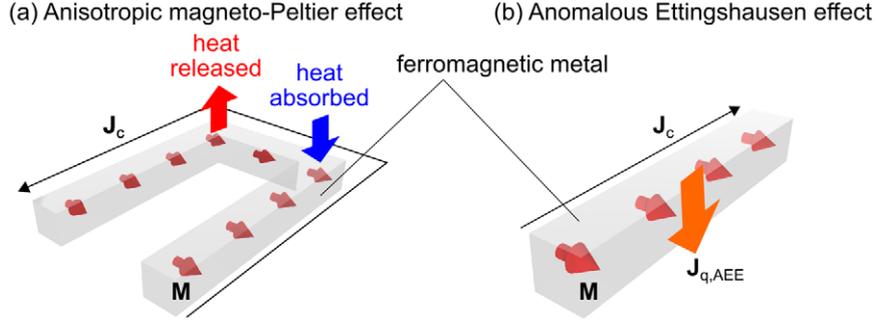

FIG. 1. Schematics of (a) the anisotropic magneto-Peltier effect (AMPE) and (b) the anomalous Ettingshausen effect (AEE) in a ferromagnetic metal. $\mathbf{J_c}$, $\mathbf{J_{q,AEE}}$, and $\mathbf{M}$ denote the charge current, heat current driven by the AEE, and magnetization vector, respectively.

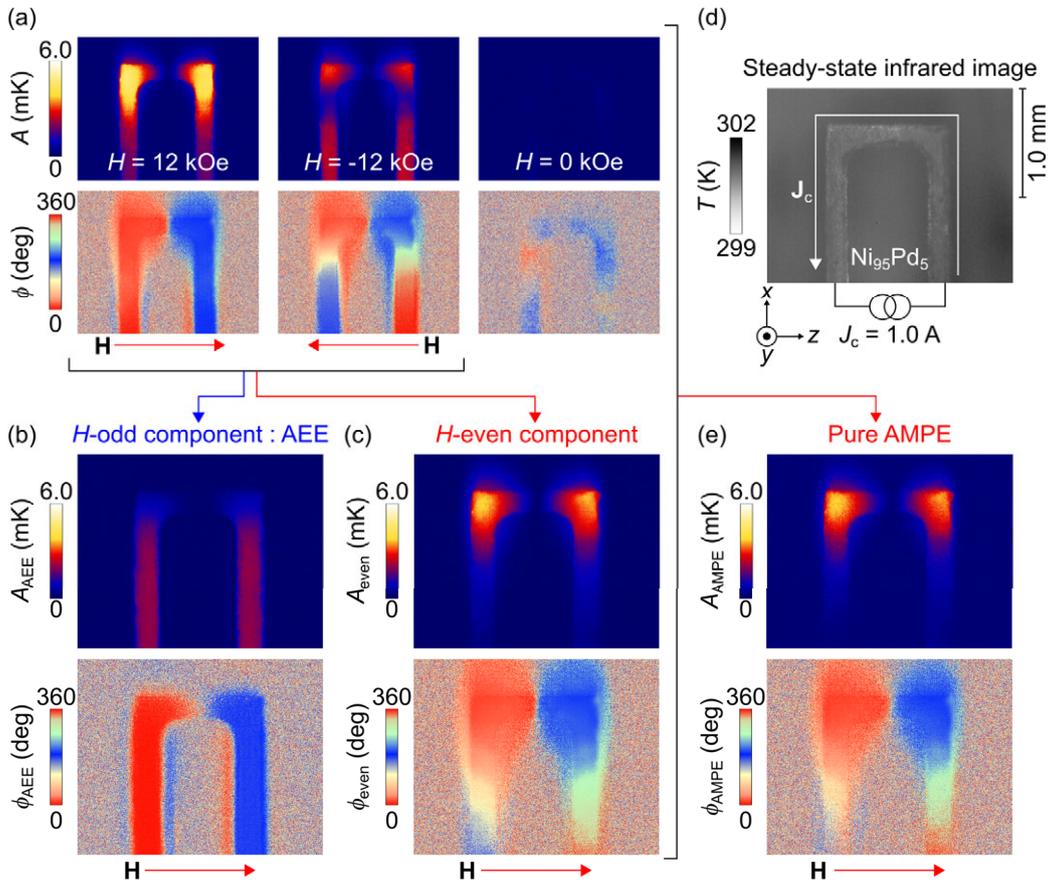

FIG. 2. (a) Lock-in amplitude $A$ and phase $\phi$ images at room temperature for the U-shaped Ni$_{95}$Pd$_5$ slab at the lock-in frequency $f = 25$ Hz and square-wave-modulated AC charge current amplitude $J_c = 1.0$ A for various values of the external magnetic field $H$. $\mathbf{H}$ denotes the direction of $H$, which is along the $z$ direction. (b) $A_\text{AEE}$ and $\phi_\text{AEE}$ images for the U-shaped Ni$_{95}$Pd$_5$ slab at $f = 25$ Hz, $J_c = 1.0$ A, and $|H| = 12$ kOe. (c) $A_\text{even}$ and $\phi_\text{even}$ images. $A_\text{AEE(even)}$ and $\phi_\text{AEE(even)}$ are the $H$-odd ($H$-even) components of the lock-in amplitude and phase, respectively, where the $H$-odd component corresponds to the AEE contribution. (d) A steady-state infrared image for U-shaped Ni$_{95}$Pd$_5$ slab at room temperature. (e) $A_\text{AMPE}$ and $\phi_\text{AMPE}$ images for the U-shaped Ni$_{95}$Pd$_5$ slab at $f = 25$ Hz, $J_c = 1.0$ A, and $|H| = 12$ kOe. $A_\text{AMPE}$ and $\phi_\text{AMPE}$ show the temperature modulation due purely to the AMPE.

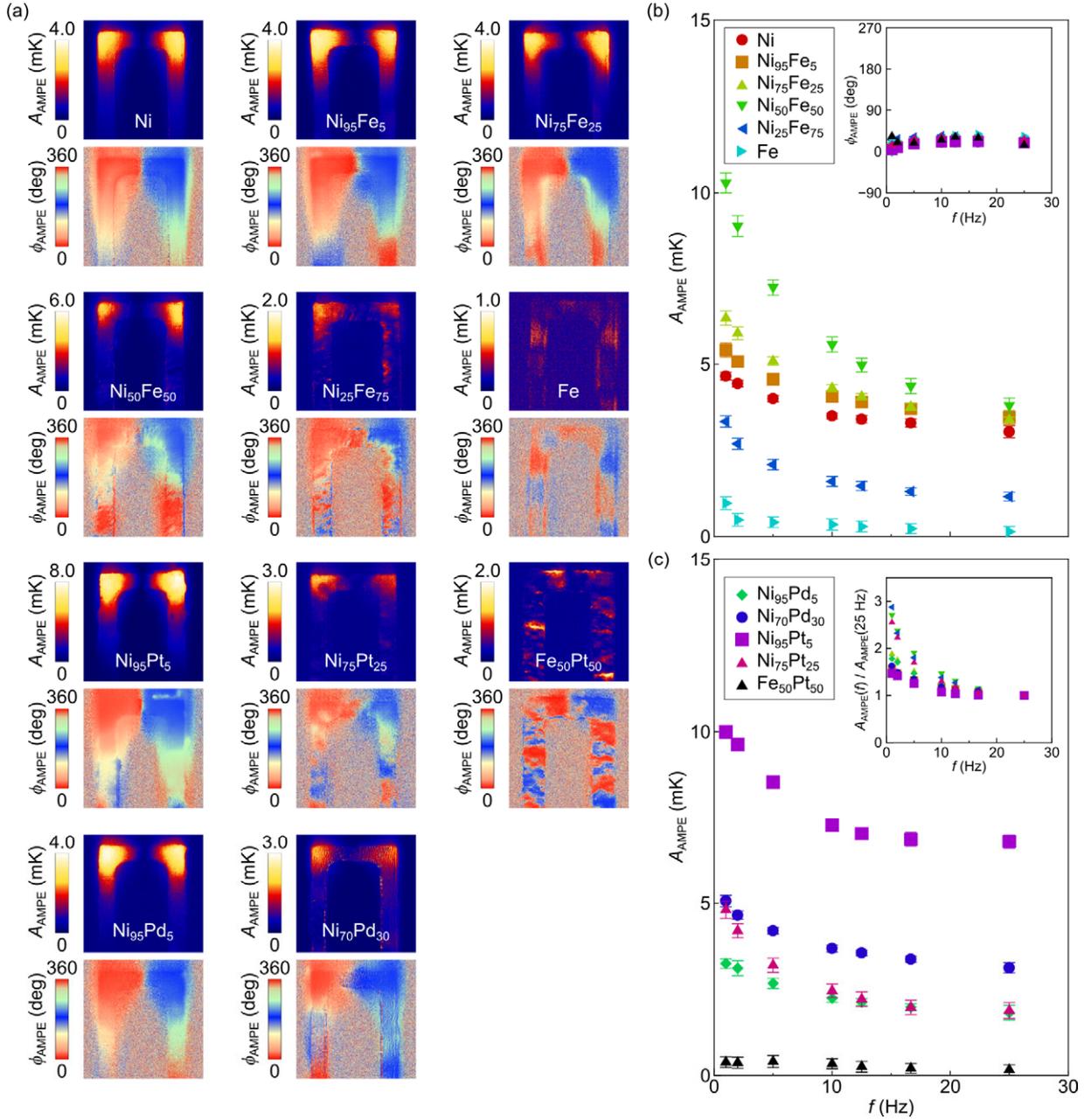

FIG. 3. (a) $A_\text{AMPE}$ and $\phi_\text{AMPE}$ images at room temperature for the U-shaped Ni, Fe, Ni-Fe alloys (Ni$_{95}$Fe$_5$, Ni$_{75}$Fe$_{25}$, Ni$_{50}$Fe$_{50}$, and Ni$_{25}$Fe$_{75}$), Ni-Pd alloys (Ni$_{95}$Pd$_5$ and Ni$_{70}$Pd$_{30}$), Ni-Pt alloys (Ni$_{95}$Pt$_5$ and Ni$_{75}$Pt$_{25}$), and Fe-Pt alloy (Fe$_{50}$Pt$_{50}$) at $f$ = 25 Hz, $J_\text{c}$ = 1.0 A, and $|H|$ = 12 kOe. (b),(c) $f$ dependence of $A_\text{AMPE}$ at the left corner of the U-shaped ferromagnets at $J_\text{c}$ = 1.0 A and $|H|$ = 12 kOe. The inset to (b) shows the $f$ dependence of $\phi_\text{AMPE}$. The inset to (c) shows the $f$ dependence of $A_\text{AMPE}(f)/A_\text{AMPE}(f = 25\text{ Hz})$.

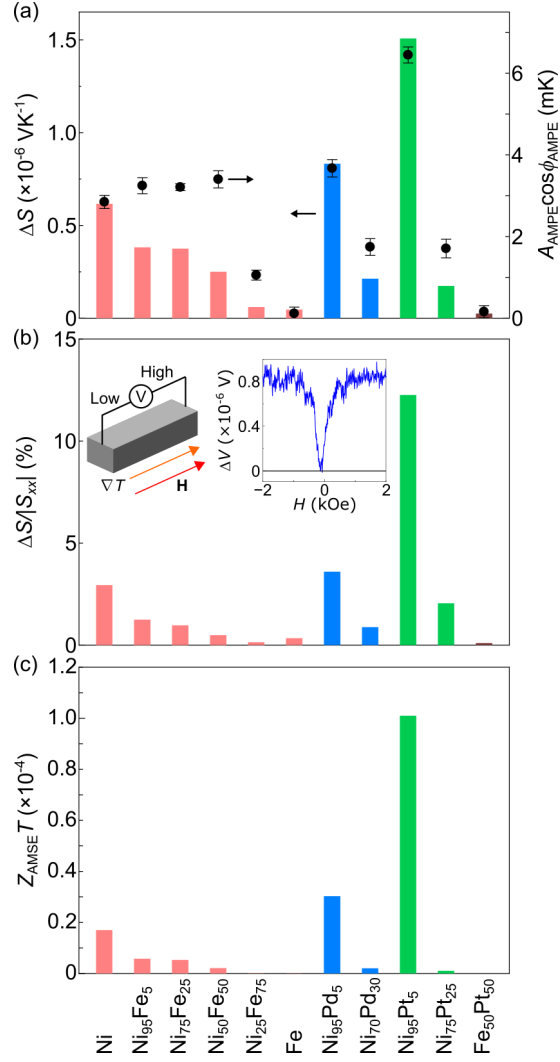

FIG. 4. (a) Anisotropy of the magneto-Seebeck coefficient $\Delta S$, obtained from the thermoelectric voltage measurements using the bar-shaped ferromagnets, and the $A_{\text{AMPE}}\cos\phi_{\text{AMPE}}$ values at the left corner of the U-shaped ferromagnets, obtained from the lock-in thermography measurements at $f = 25$ Hz, $J_c = 1.0$ A, and $|H| = 12$ kOe, at room temperature. (b),(c) $\Delta S/|S_{xx}|$ and $Z_{\text{AMSE}}T$ at the temperature of $T = 300$ K for the ferromagnets. $S_{xx}$ and $Z_{\text{AMSE}}T$ respectively denote the Seebeck coefficient and the dimension-less figure of merit for the anisotropic magneto-Seebeck effect (AMSE), defined by Eq. (3). The inset to (b) shows a schematic of the AMSE-measurement configuration and the $H$ dependence of the AMSE-induced voltage $\Delta V$. After measuring the $H$-$\Delta V$ curves, the temperature difference between the voltage probes and the resultant Seebeck coefficient were calibrated by using the Seebeck Coefficient/Electric Resistance Measurement System (ZEM-3, ADVANCE RIKO, Inc.).

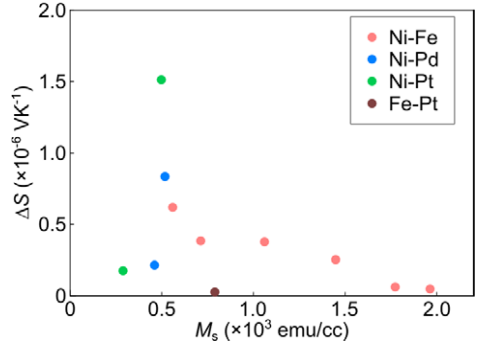

FIG. 5. Saturation magnetization $M_s$ dependence of $\Delta S$ for the Ni-Fe (pink), Ni-Pd (blue), Ni-Pt (green), and Fe-Pt (brown) alloys. Pure Ni and Fe are included in the Ni-Fe alloys.

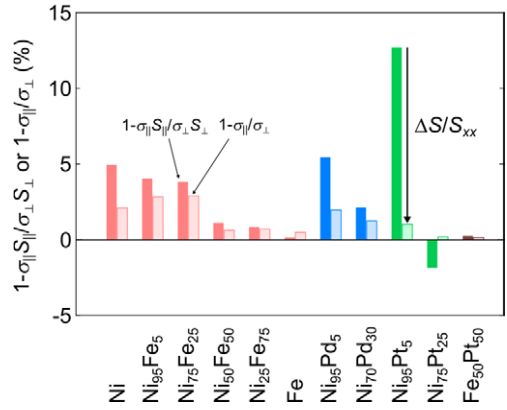

FIG. 6. $1 - \sigma_\parallel/\sigma_\perp$ and $1 - \sigma_\parallel S_\parallel/(\sigma_\perp S_\perp)$ values of the ferromagnets. $\sigma_\parallel$ ($\sigma_\perp$) and $S_\parallel$ ($S_\perp$) are the electrical conductivity and the Seebeck coefficient for the $\mathbf{M} \parallel \nabla T$ ($\mathbf{M} \perp \nabla T$) configuration, respectively.

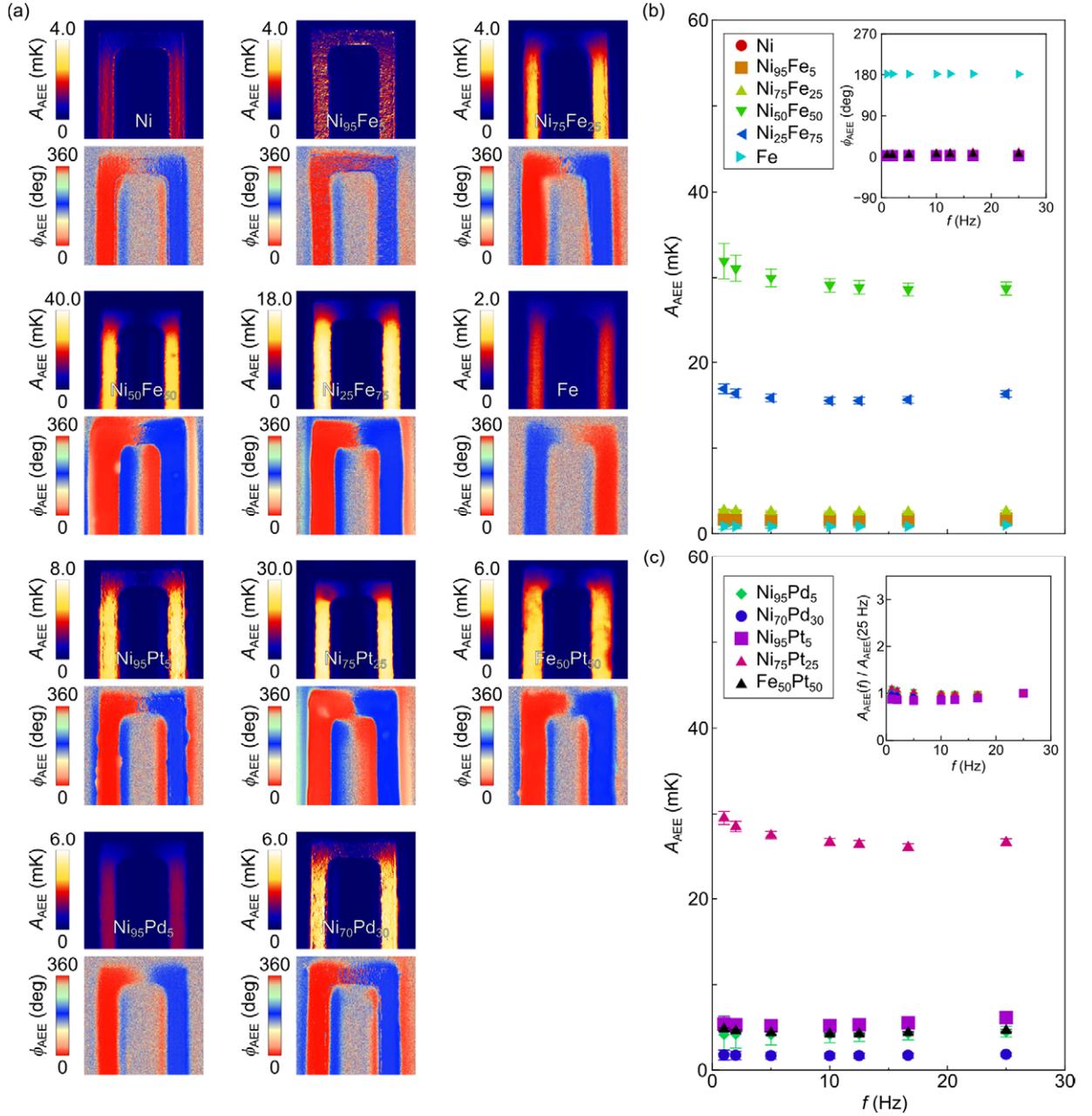

FIG. 7. (a) $A_{\text{AEE}}$ and $\phi_{\text{AEE}}$ images at room temperature for the U-shaped Ni, Fe, Ni-Fe alloys (Ni$_{95}$Fe$_5$, Ni$_{75}$Fe$_{25}$, Ni$_{50}$Fe$_{50}$, and Ni$_{25}$Fe$_{75}$), Ni-Pd alloys (Ni$_{95}$Pd$_5$ and Ni$_{70}$Pd$_{30}$), Ni-Pt alloys (Ni$_{95}$Pt$_5$ and Ni$_{75}$Pt$_{25}$), and Fe-Pt alloy (Fe$_{50}$Pt$_{50}$) at $f$ = 25 Hz, $J_c$ = 1.0 A, and $|H|$ = 12 kOe. (b),(c) $f$ dependence of $A_{\text{AEE}}$ at the left corner of the U-shaped ferromagnets at $J_c$ = 1.0 A and $|H|$ = 12 kOe. The inset to (b) shows the $f$ dependence of $\phi_{\text{AEE}}$. The inset to (c) shows the $f$ dependence of $A_{\text{AEE}}(f)/A_{\text{AEE}}(f = 25\text{ Hz})$.

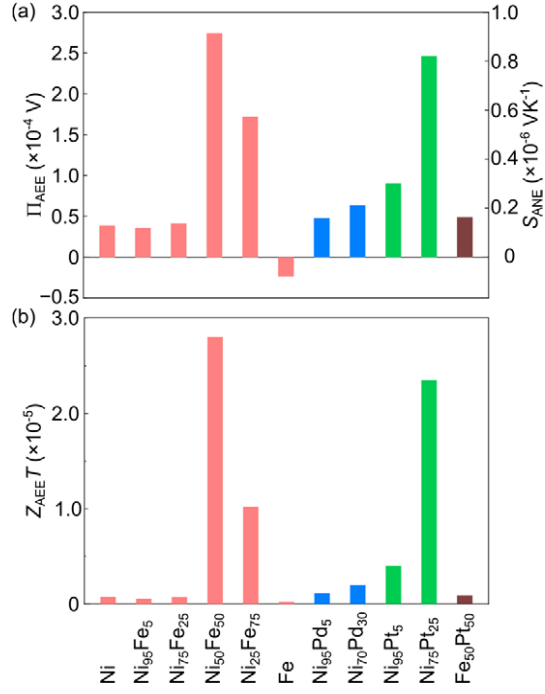

FIG. 8. (a) Anomalous Ettingshausen coefficient $\Pi_{AEE}$ and corresponding anomalous Nernst coefficient $S_{ANE}$ at $T = 300$ K of the ferromagnets. (b) Dimension-less figure of merit for the AEE $Z_{AEE}T$ at $T = 300$ K of the ferromagnets.

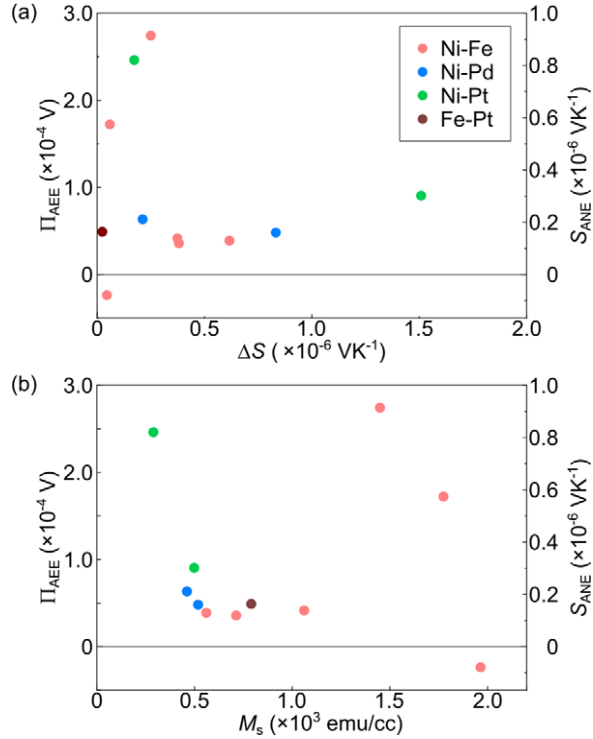

FIG. 9. (a) $\Delta S$ dependence of $\Pi_{AEE}$ and $S_{ANE}$ for the Ni-Fe (pink), Ni-Pd (blue), Ni-Pt (green), and Fe-Pt (brown) alloys. Pure Ni and Fe are included in the Ni-Fe alloys. (b) $M_s$ dependence of $\Pi_{AEE}$ and $S_{ANE}$ for the alloys.

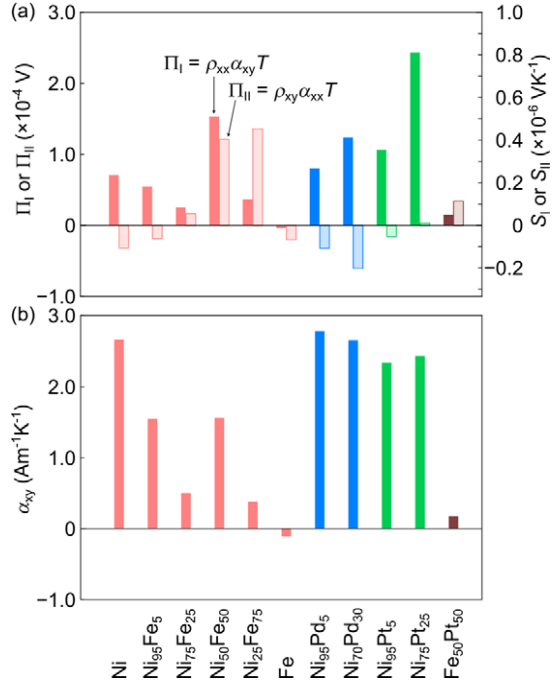

FIG. 10. (a) $\Pi_{\rm I}$ and $\Pi_{\rm II}$ defined by Eq. (9) of the ferromagnets. The corresponding $S_{\rm I}$ and $S_{\rm II}$ values at $T = 300$ K are also shown. (b) Transverse thermoelectric conductivity $\alpha_{xy}$ of the ferromagnets.

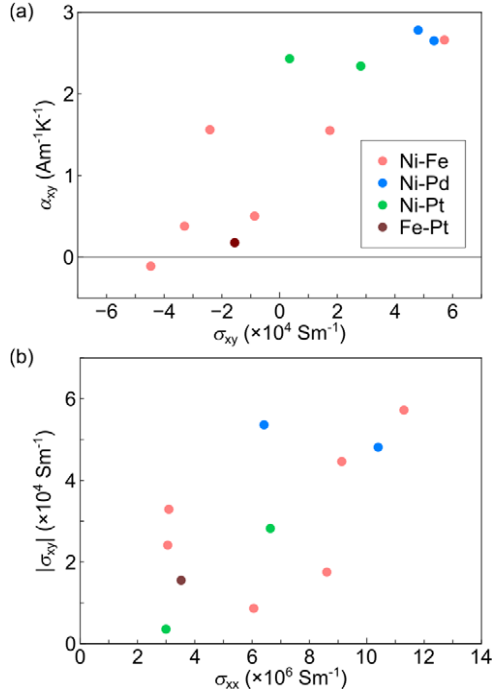

FIG. 11. (a) Anomalous Hall conductivity $\sigma_{xy}$ dependence of $\alpha_{xy}$ for the Ni-Fe (pink), Ni-Pd (blue), Ni-Pt (green), and Fe-Pt (brown) alloys. (b) Longitudinal conductivity $\sigma_{xx}$ dependence of $|\sigma_{xy}|$ for the ferromagnets.

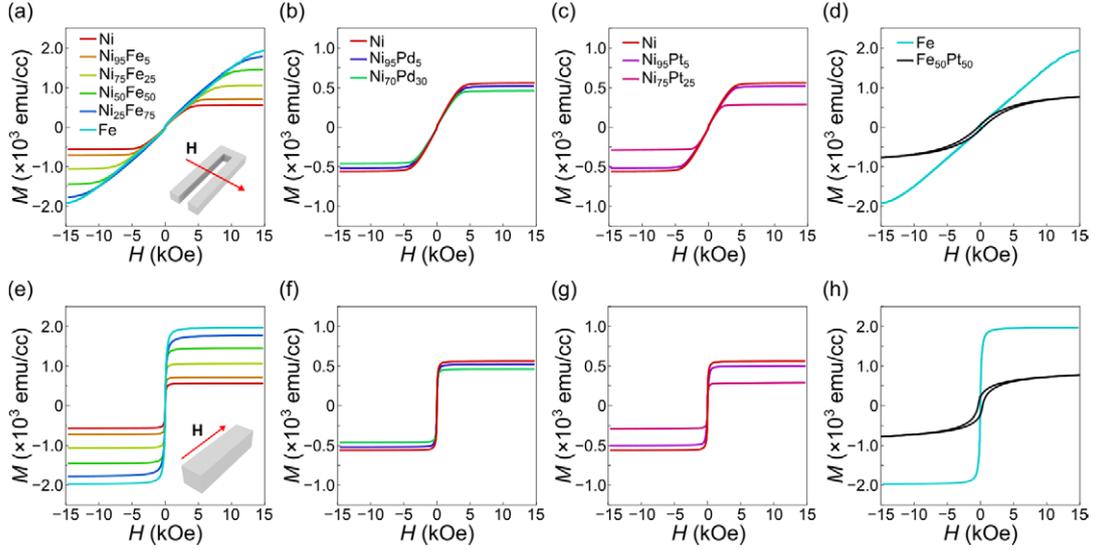

FIG. 12. Magnetization $M$ curve of (a)-(d) the U-shaped ferromagnets and (e)-(h) the bar-shaped ferromagnets. The **H** directions during the measurements of the $M$-$H$ curves are schematically depicted in the insets to (a) and (e).

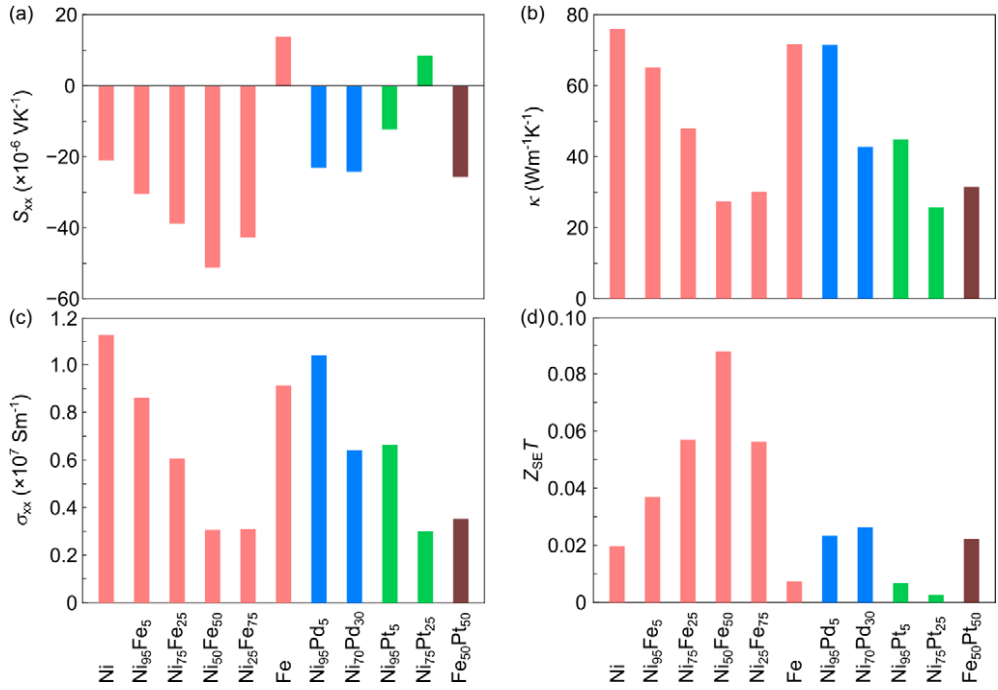

FIG. 13. (a) $S_{xx}$, (b) thermal conductivity $\kappa$, (c) $\sigma_{xx}$, and (d) dimension-less figure of merit for the Seebeck effect $Z_{SE}T$ at $T = 300$ K and $H = 0$ kOe for the ferromagnets.

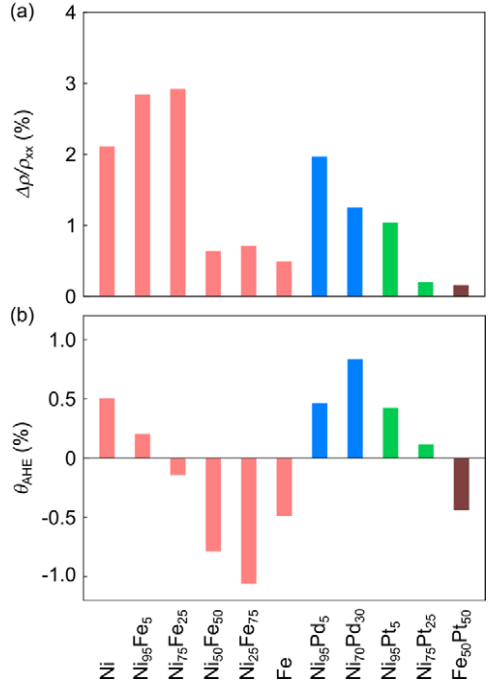

FIG. 14. (a) Anisotropic magnetoresistance ratio $\Delta\rho/\rho_{xx}$ and (b) anomalous Hall angle $\theta_{AHE}$ for the ferromagnets. $\rho_{xx}$ is the longitudinal electrical resistivity at $H = 0$ kOe. $\Delta\rho$ is defined as $\rho_\parallel - \rho_\perp$ with $\rho_\parallel$ ($\rho_\perp$) being the electrical conductivity for the $\mathbf{M} \parallel \mathbf{J}_c$ ($\mathbf{M} \perp \mathbf{J}_c$) configuration, and estimated by substituting the measured values of $\rho_\parallel$ and $\rho_{xx}$ into the relation $\Delta\rho = 3(\rho_\parallel - \rho_{xx})/2$ for isotropic ferromagnets [43]. $\theta_{AHE}$ is defined as $\rho_{xy}/\rho_{xx}$ with being $\rho_{xy}$ is the off-diagonal component of the electrical resistivity tensor.